# Detuning Tunable OAM Generation via Double-Λ Four-Wave Mixing in Hot Rubidium Vapor

Shahar Monsa, Michael Shulinder, Shmuel Sternklar, Eliran Talker*

Department of Electrical and Electronic Engineering, Ariel University, Ariel 40700, Israel

*Corresponding author: elirant@ariel.ac.il

**Abstract**

We demonstrate detuning-tunable generation of orbital-angular-momentum (OAM) light using a double-Λ four-wave-mixing (FWM) process in Doppler-broadened rubidium vapor. Two near-resonant pumps on the D1 line drive non-degenerate FWM that produces bright probe–conjugate beams whose transverse modes evolve with pump detuning. A paraxial density-matrix model coupled to split-step propagation predicts detuning-dependent spatial gain shaping that sets the OAM content; experiments with a mode-cleaned pump laser and a 12-mm AR-coated vapor cell validate these predictions. We quantify mode formation by imaging, spectroscopy, and power measurements, and verify OAM conservation between the generated beams. The results establish resonant atomic vapor as a compact, tunable platform for structured-light generation with applications to high-dimensional quantum communications and imaging.

**Introduction**

The study of the orbital angular momentum (OAM) of light, electrons[1–4], and neutrons[5–7] has emerged as a rapidly evolving frontier in modern physics. In optics, the transverse spatial structure of light, and particularly its OAM, can be utilized in a broad range of applications, from optical tweezing[8–10] and manipulation of dielectric particles[11–13], micro-machines, and biological specimens[14–16] to the guiding and rotation of ultracold atoms[17–19]. In addition, this technology has the potential for dramatically increasing the information capacity of both classical and quantum communication systems. To realize this potential, advanced methods for generating, manipulating, and frequency-converting OAM states are indispensable.

OAM originates from two distinct components: spin angular momentum, associated with polarization[20], and the dependence on the spatial distribution of the light field. Allen *et al.*[21] showed that paraxial Laguerre–Gaussian beams carry well-defined

OAM along the propagation direction, quantified by the integer topological charge $\ell$. Manipulation of OAM beams has since been explored in both linear and nonlinear regimes. In nonlinear optics, these beams have been employed in second-harmonic generation (SHG) in crystals, where a fundamental beam carrying charge l produces a frequency-doubled beam with charge $2\ell$, reflecting conservation of OAM. Such processes highlight a central principle: the transfer of topological charge and transverse structure from the driving to the generated field. Today, SHG and other second-order processes provide versatile platforms for probing the transverse degrees of freedom of light.

Correlations in OAM have also been studied in spontaneous parametric down-conversion (SPDC), where spatial entanglement between signal and idler photons reflects the OAM of the pump[22]. For higher-order effects, four-wave mixing (FWM) constitutes a powerful third-order nonlinear process for OAM generation. A particularly fruitful configuration involves FWM in hot rubidium vapor, driven by amplified spontaneous emission in a cascaded scheme. Here, photons at 776 nm and 780 nm combine to generate blue light at 420 nm along with mid-infrared emission at 5.23 μm, enabling direct transfer of OAM from the near-infrared pumps to the generated fields, and supporting the creation of complex transverse beam structures[23–25].

In this work, we investigate OAM generation at different frequency detuning using a non-degenerate double-Λ FWM configuration in hot rubidium vapor[26–29]. Two optical fields near 795 nm interact with the atomic ensemble to produce bright probe and conjugate beams. As a parametric process, the double-Λ scheme ensures that the atomic system returns to its initial state after interaction, producing probe and conjugating photons in pairs when FWM dominates. Phase-matched nonlinear interactions maintain coherence, both longitudinal and transverse, between the driving and generated fields, while guaranteeing OAM conservation via the spiral phase front. This framework provides precise control of transverse optical modes. Our results advance the understanding of OAM generation and manipulation in resonant atomic media, offering new routes for applications in quantum information processing, high-dimensional entanglement, and quantum communication.

**Theoretical model:**

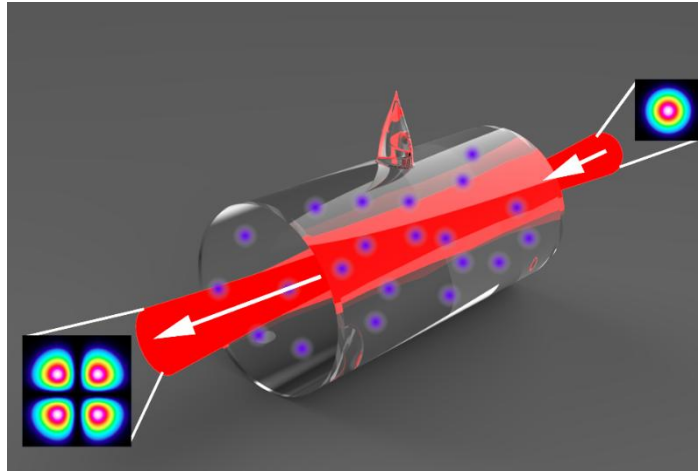

**Figure 1.** Conceptual illustration of detuning-tunable OAM generation in a hot rubidium vapor cell via double-Λ four-wave mixing. A near-Gaussian pump propagates through the vapor, where spatial gain shaping and phase matching convert the transverse mode of the generated light from a multi-lobed pattern to a vortex profile. Total OAM is conserved between the probe and conjugate beams (not shown separately).

We consider an ensemble of Doppler-broadened rubidium atoms, each atom modeled as a four-level system in a double-Λ spontaneous FWM configuration, as illustrated in Figure 1(b). The electric dipole-allowed transitions |1⟩↔|3⟩ and |2⟩↔|4⟩ are driven by two strong pump fields with Rabi frequency Ω. The remaining transitions, |4⟩↔|1⟩ and |3⟩↔|2⟩, correspond to the generated probe ($\Omega_{pr}$) and conjugate ($\Omega_{conj}$) fields, respectively. The atomic dynamics are described by the standard density matrix formalism, from which the equations of motion are derived[30,31].

(1) $$\frac{\partial \rho}{\partial t} = -\frac{i}{\hbar}[\widehat{\mathcal{H}}, \rho] + \mathcal{L}\rho$$

Where $\mathcal{L}\rho$ is Lindblad operator which describes population and coherence time decay. It is equal to

(2) $$\mathcal{L}\rho = \sum_d \frac{\Gamma_d}{2}\left(\sigma_d^\dagger \sigma_d \rho + \rho \sigma_d^\dagger \sigma_d - 2\sigma_d \rho \sigma_d^\dagger\right)$$

where $\sigma_d^\dagger = |n\rangle\langle m|$ and $\Gamma_d$ is the decay rate from state $n, m \in \{|1\rangle, |2\rangle, |3\rangle, |4\rangle\}$.

The Hamiltonian of the atoms combined with the Hamiltonian of the light-atom interaction is given by

(3) $$\widehat{\mathcal{H}} = \widehat{\mathcal{H}}_{atoms} + \widehat{H}_I$$

(4) $$\hat{\mathcal{H}}_{atoms} = \hbar\omega_{HFS}|2\rangle\langle 2| + \hbar\Delta_2|3\rangle\langle 3| + \hbar(\Delta_1 + \omega_{HFS})|4\rangle\langle 4|$$

(5) $$\hat{\mathcal{H}}_I = -\hbar\Big(\Omega e^{i(k_{pu}\cdot r-\omega_{pu}t)}|1\rangle\langle 3| + \Omega e^{i(k_{pu}\cdot r-\omega_{pu}t)}|2\rangle\langle 4|$$
$$+ \Omega_{pr}e^{i(k_{pr}\cdot r-\omega_{pr}t)}|1\rangle\langle 4| + \Omega_{conj}e^{i(k_{conj}\cdot r-\omega_{conj}t)}|2\rangle\langle 3|\Big)$$

where $\omega_{HFS} = 2\pi \cdot 3.035$ GHz for $^{85}$Rb and $\omega_{HFS} = 2\pi \cdot 6.8$ GHz for $^{87}$Rb. $\Delta_1$ is the detuning from the resonance transition between level $|2\rangle$ to level $|4\rangle$. $\Delta_2$ is the detuning from the resonance transition between level $|1\rangle$ to level $|3\rangle$. $\omega_{pu}, \omega_{pr}, \omega_{conj}$ are the frequencies for the pump, probe, and conjugate beams respectively. $k_{pu}, k_{pr}, k_{conj}$ are the wave vectors of the pump, probe and conjugate beams respectively.

The induced macroscopic polarization can be expressed in terms of the atomic coherence as

(20) $$P_{pr} = \mathcal{N}\big(d_{14}\rho_{41}e^{-i\omega_{pr}t} + c.c.\big)$$
(21) $$P_{conj} = \mathcal{N}\big(d_{23}\rho_{32}e^{-i\omega_{conj}t} + c.c.\big)$$

Here, $\mathcal{N}$ is the atomic density, $d_{23}$ is the dipole moment and $\rho_{ij}$ is the atomic operators under the Heisenberg model (see the Appendix). Equations (20) and (21) can be recast as

(22) $$P_{pr} = \varepsilon_0\chi_{pp}(\omega_{pr})E_{pr}e^{ik_{pr}r} + \varepsilon_0\chi_{pc}(\omega_p)E^*_{conj}e^{i(2k_0-k_{conj})}$$
(23) $$P_{conj} = \varepsilon_0\chi_{cc}(\omega_{conj})E_{conj}e^{ik_{conj}r} + \varepsilon_0\chi_{cp}(\omega_p)E^*_{pr}e^{i(2k_0-k_{pr})}$$

Here, the two coefficients $\chi_{pp}$ and $\chi_{cc}$ (see the Appendix) describes the effective linear polarization process for the probe and the conjugate fields, respectively, and unlike the usual linear coefficients, they depend nonlinearly on the pump field. The other coefficients $\chi_{pc}$ and $\chi_{cp}$ are responsible for the four-wave mixing process. $k_{pr}, k_{conj}$ and $k_0$ are the wave number for the probe, conjugate and pump beam respectively. The wave equation for the generated probe and conjugate light beams can be written as

(24) $$\left(\nabla^2 + \frac{1}{c^2}\frac{\partial^2}{\partial t^2}\right)E_{conj,pr} = \frac{4\pi}{c^2}\frac{\partial^2 P_{conj,pr}}{\partial t^2}$$

where $E_{conj,pr}$ is the electric field of the generated probe and conjugate signal and $P_{conj,pr}$ is the induced polarization for the conjugate and the probe field.

Under slowly varying envelope and paraxial wave approximation, the wave equation can be cast into the following form

(25) $$\frac{\partial \Omega_{pr}}{\partial z} = \frac{i}{2k_{pr}} \nabla_\perp^2 \Omega_{pr} + i\eta \langle \rho_{41} \rangle$$

(26) $$\frac{\partial \Omega_{conj}}{\partial z} = \frac{i}{2k_{conj}} \nabla_\perp^2 \Omega_{conj} + i\eta \langle \rho_{32} \rangle$$

where $\Omega_{pr} = d_{41} \cdot E_{pr}/\hbar$ and $\Omega_{conj} = d_{32} \cdot E_{conj}/\hbar$. The derivative on the left-hand side indicates the variation of the amplitude of the generated signal envelope, $\Omega_{pr}, \Omega_{conj}$ along the length of the medium. The first term on the right-hand side represents the beam's phase induced diffraction and the rotation of the wave front during its propagation. The last term on the right-hand side accounts for the generation and dispersion of the medium. The coupling constant $\eta$ is defined as $\eta = 3\mathcal{N}\gamma\lambda^2/8\pi$ where $\mathcal{N}$ is the atomic density, and $\lambda$ is the wavelength of the probe or conjugate beam. We use appropriate spatially dependent transverse profile of the optical fields to generate the spontaneous FWM signal. For this purpose, the transverse spatial profile of the optical beams is taken to be of $LG_p^\ell$ with radial index $p$ and topological charge $\ell$

(27) $$E_{pump}(r,\phi,z) = E_{pump}^0 \frac{w_j}{w(z)} \left(\frac{r\sqrt{2}}{w(z)}\right)^{|\ell|} e^{-\frac{r^2}{w^2(z)}} L_m^\ell \left[\frac{2r^2}{w^2(z)}\right]$$
$$\cdot e^{i\ell\phi} e^{\frac{ik_0 r^2}{2R(z)}} e^{-i(2m+|\ell|+1)\tan^{-1}\left(\frac{z}{z_0}\right)} = E_{pump}^0 \cdot u_{\ell,p}(r,\phi,z)$$

where $r = \sqrt{x^2 + y^2}$ and $\phi = \tan^{-1}\left(\frac{y}{x}\right)$. $\Omega_j^0$ is the input amplitude, $R(z) = z + z_0^2/z$ is the radius of curvature and $z_0 = \pi\omega_j^2/\lambda$ is the Rayleigh length of the beam. The beam spot size is defined as $\omega_{pump}(z) = \omega_j\sqrt{1 + (z/z_0)^2}$. By inserting equation (24) into equations (22) and (23) we can calculate both the probe and the conjugates fields after exiting from the vapor cell. The combinations of OAM in two degenerate four-wave mixing processes in Rb vapor can be found by solving equations (22)-(23) at the exit of the rubidium cell. It can be expressed as a superposition of LG modes:

(28) $$E_{pr}(r,\phi,z) = \sum_{l,p} a_{\ell,p} u_{\ell,p}(r,\phi,z)$$

The $a_{l,p}$ coefficients are equal to

(29) $$a_{\ell,p} = \kappa_1 \int_{-L/2}^{L/2} \left( \int\int E_{pu}^2 E_{pr}^* u_{\ell,p}^*(r,\phi,z) d^2 r_\perp \right) e^{-i\Delta k z} dz$$

$$= \kappa_1 \int_{-L/2}^{L/2} \Lambda_p^\ell(z) e^{-i\Delta k z} dz$$

Here $\Delta k = 2k_0 - k_{pr} - k_{conj}$ where $k_0$ is the wavenumber of the pump beam. In a thin medium, where $L/z_0 \ll 1$, we can approximate

(30) $$a_{l,p} \approx \kappa_1 T_1(L) \Lambda_p^\ell(0)$$

where $T_1(L) = \int_{-L/2}^{L/2} e^{-i\Delta k z} dz = 2\sin\left(\frac{\Delta k_1 L}{2}\right)$.

A straightforward solution for the $E_{conj}$ can be found by simply replacing $E_{pr}$ by $E_{conj}$. For $E_{pr}^0 = u_{l_{pr},0}$ and $E_{conj}^0 = u_{l_{conj},0}$ the generated fields at the detection position can be expressed as

(31) $$E_{pr}(r,\phi,z_d) = e^{i(2\ell_{pr}-\ell_{conj})\phi} \mathcal{A}_1(r,z_d)$$
(32) $$E_{pr}(r,\phi,z_d) = e^{i(2\ell_{conj}-\ell_{pr})\phi} \mathcal{A}_2(r,z_d)$$

where the functions $\mathcal{A}_{1,2}(r,z_d)$ describe the radial amplitude distributions and phase profiles, $\mathcal{A}_{1,2}(r,z_d) = \sum_p \alpha_{l_{1,2},p} V_p^{|\ell|}(r,z)$.

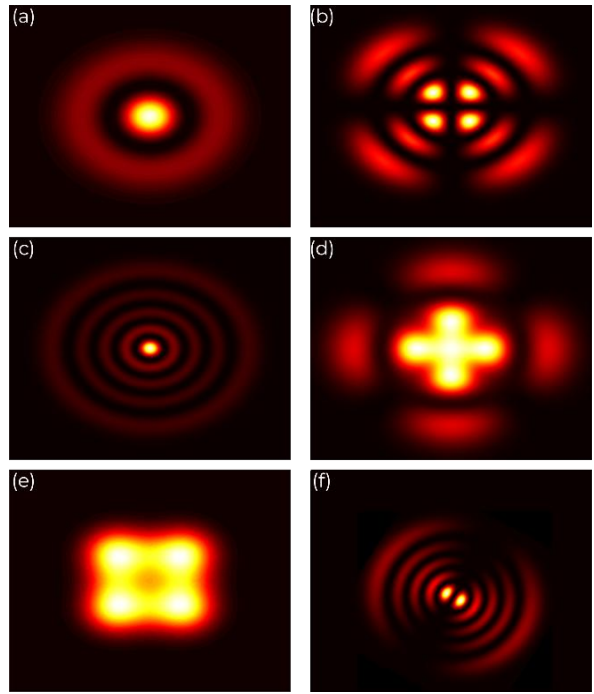

**Figure 2**. Intensity pattern of the Doppler-averaged atomic coherence $\langle \rho_{coh} \rangle$ in the transverse plane as a function of pump detuning (a) $\Delta_2 = 3$ GHz blue detune from $^{87}$Rb $F = 1 \rightarrow F'$ transition (b) $\Delta_2 = 3$

GHz red detune from $^{87}$Rb $F = 2 \to F'$ transition (c) $\Delta_2 = 1.5$ GHz blue detune from $^{87}$Rb $F = 2 \to F'$ transition (d) $\Delta_2 = 1.5$ GHz blue detune from $^{87}$Rb $F = 1 \to F'$ transition (e) $\Delta_2 = 1.5$ GHz red detune from $^{85}$Rb $F = 3 \to F'$ transition (f) $\Delta_2 = 3$ GHz red detune from $^{87}$Rb $F = 2 \to F'$ transition. The parameter are m=0, l=0, $\omega_1 = 120\ \mu m$, $\Omega = 20 \cdot \Gamma$, $T = 394\ K$

OAM modes were calculated by numerically solving equations 31-32 using a split-step Fourier method (SSFM)[32] to incorporate diffraction, combined with a fourth-order Runge–Kutta integration for the source term. The induced atomic coherence is defined as $\langle\rho_{co}\rangle = \langle\rho_{41}\rangle + \langle\rho_{32}\rangle$, reflecting the equal contributions of probe and conjugate channels in the double-Λ FWM configuration, where the parametric nature of the process ensures identical initial and final atomic states. Accordingly, the number of photons added to the probe and conjugate beams is equal, provided FWM is the dominant nonlinear interaction in the medium. The Doppler-averaged nonlinear coherence $\langle\rho_{co}\rangle$ governs the observed OAM conversion processes. A pump beam with Rabi frequency $\Omega = 20\Gamma$ ($\Gamma = 2\pi \times 5.75$ MHz) was employed, with detuning values chosen to satisfy the double-Λ resonance condition, $\Delta_1 = \Delta_2 + \omega_{HFS}$. The pump was prepared in the fundamental Laguerre–Gaussian mode (LG$_{00}$) prior to entering the vapor cell, leading to the generation of the spatially structured OAM modes. Figure 2 shows the numerical simulations of the total output field, obtained from the coherent sum of probe and conjugate contributions, for different pump detuning. Each panel (a–f) illustrates the transverse intensity distribution of the generated modes, highlighting the spatial structures arising from the OAM conversion processes under varying resonance conditions.

## **Experimental setup and results:**

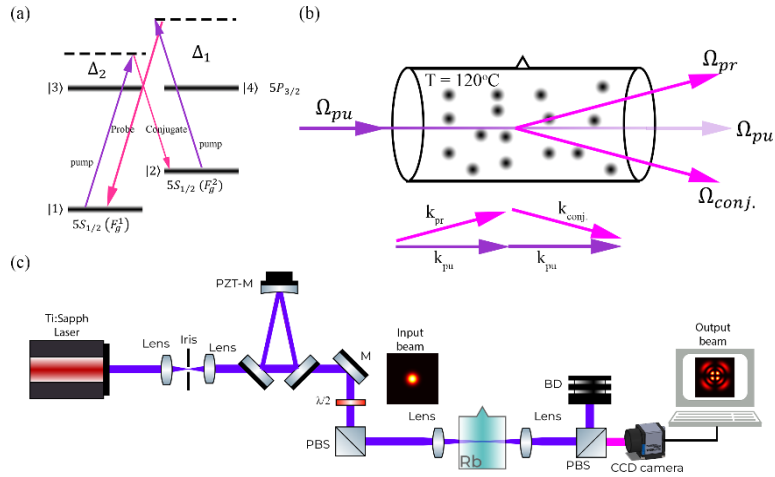

**Figure 3**. (a) schematic representation of the four-level double-Λ atomic system. The energy state of $^{85}$Rb are defined as $|1\rangle = 5S_{1/2}\left(F_g^1\right)$ where $F_g^1 = 2$ for $^{85}$Rb and $F_g^1 = 1$ for $^{87}$Rb, $|1\rangle = 5S_{1/2}\left(F = 3\right)$ where $F_g^2 = 3$ for $^{85}$Rb and $F_g^2 = 2$ for $^{87}$Rb, $|3\rangle$, $|4\rangle = 5P_{3/2}$. (b) a simple block diagram of the model system and the phase matching condition (c) The experimental setup used for the generation of OAM modes. M: mirror; PBS: polarizing beam splitter; BD: beam dump; CCD (charge-coupled device) camera

The generation of OAM modes via FWM in rubidium vapor is shown schematically in Figure 3. A 500 mW pump beam from a Ti:Sapphire laser was focused to a waist of 120 μm (1/e² intensity radius), corresponding to a Rabi frequency of $\Omega \approx 20\Gamma$. The laser was tuned across the D$_1$ transition of rubidium (λ = 795 nm). To suppress beam jitter and higher-order spatial modes, the pump was coupled into a high-finesse three-mirror ring cavity acting as a mode cleaner, which resonantly enhanced only the LG$_{00}$ mode while rejecting higher-order contributions. Additional spatial filtering was achieved using a lens–iris system. The output was then passed through a half-wave plate and a polarizing beam splitter (PBS) to control the pump power and ensure high-purity linear polarization before entering a 12 mm-long anti-reflection–coated vapor cell contains natural abundance rubidium, held at ~120 °C, corresponding to an atomic density of N = 2.6 × 10$^{12}$ cm$^{-3}$. After exiting the cell, the pump was separated from the generated fields using a second PBS, and the probe and conjugate beams were imaged onto a CCD camera.

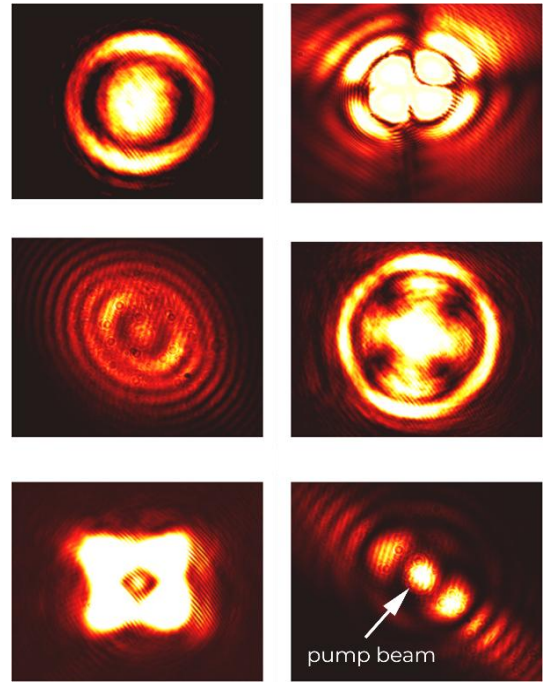

**Figure 4**. Measured intensity pattern of the OAM for different pump beam detuning (a) $\Delta_2 = 3$ GHz blue detune from $^{87}$Rb $F = 1 \rightarrow F'$ transition (b) $\Delta_2 = 3$ GHz red detune from $^{87}$Rb $F = 2 \rightarrow F'$

transition (c) $\Delta_2 = 1.5$ GHz blue detune from $^{87}$Rb $F = 2 \rightarrow F'$ transition (d) $\Delta_2 = 1.5$ GHz blue detune from $^{87}$Rb $F = 1 \rightarrow F'$ transition (e) $\Delta_2 = 1.5$ GHz red detune from $^{85}$Rb $F = 3 \rightarrow F'$ transition (f) $\Delta_2 = 3$ GHz red detune from $^{87}$Rb $F = 2 \rightarrow F'$ transition. The cell temperature set to $120°C$ and the pump power was 500 mW. Since we are using PBS some leakage from the pump beam reaching to the CCD camera.

Figure 4 displays the experimentally measured transverse intensity distributions of OAM modes generated via double-Λ FWM in rubidium vapor. The images correspond to different pump detunings while the pump power was held constant. As the detuning is varied, the spatial structure of the output beams evolves significantly, reflecting modifications in the phase-matching conditions and the redistribution of atomic coherence within the medium. The observed mode patterns, ranging from ring-like to multi-lobed structures, directly illustrate the sensitivity of the nonlinear conversion process to frequency detuning. These measurements are in excellent agreement with the theoretical predictions of the Doppler-averaged nonlinear atomic coherence shown in Fig. 2, confirming that the coherence term ⟨ρ_co⟩ governs the OAM conversion dynamics and validates the numerical modeling of the system.

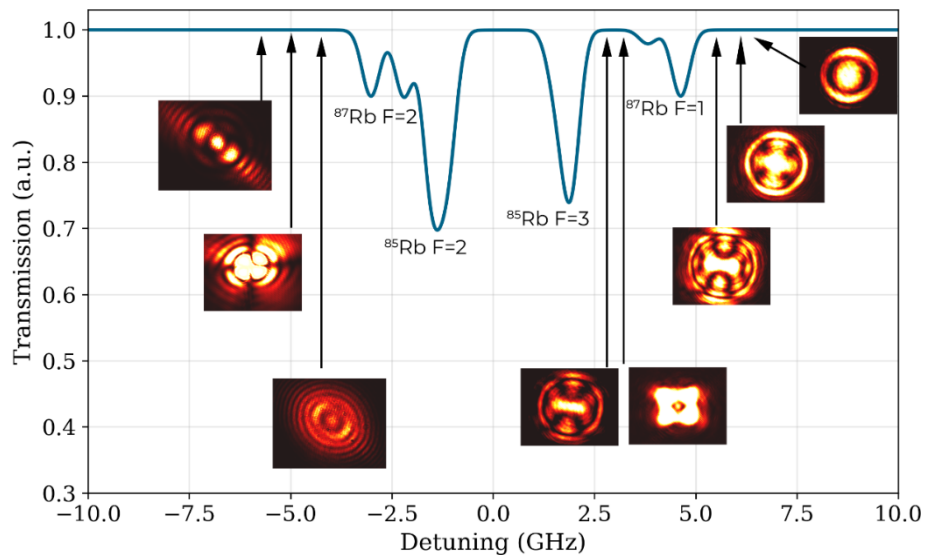

**Figure 5.** Measured intensity patterns of the OAM modes for various pump detunings across the D1 transition line.

Figure 5 shows the transmission spectrum of the rubidium D1 line together with the experimentally measured transverse intensity profiles of the generated OAM modes at selected pump detuning. The spatial patterns evolve from ring-like to multi-lobed structures as the detuning is varied, reflecting changes in phase-matching conditions and the redistribution of atomic coherence within the medium. The correspondence between the spectral position and the mode structure emphasizes the sensitivity of the FWM process to the hyperfine resonances of both isotopes. The total output power of the OAM fields generated was approximately 2 mW, the conversion efficiency with a pump power of 500 mW, the combined generated power (probe + conjugate) was 2 mW, yielding $\eta_{tot} = (P_{pr} + P_{cj})/P_{pump,in} = 0.40\%$. Assuming symmetric pair generation under double-Λ FWM, the per-beam efficiency is $\approx 0.20\%$

## Conclusion:

In summary, we have shown—both theoretically and experimentally—that double-Λ four-wave mixing in Doppler-broadened rubidium is a compact and controllable source of orbital-angular-momentum (OAM) light. With $LG_{00}$ pumps, the observed helical phase arises from detuning-dependent spatial gain shaping and transverse phase matching in the atomic medium, not from pump OAM. Varying the one-photon detuning provides a simple control knob that deterministically reshapes the transverse mode and tunes the dominant OAM content of the generated fields, in agreement with our paraxial propagation model. The measurements validate resonant hot-vapor FWM as a practical platform for structured-light generation, offering alignment tolerance, spectral tunability, and compatibility with milliwatt-level pumps. These results open a route to on-demand vortex beams for high-dimensional quantum information, multiplexed free-space, and precision metrology based on mode-selective sensing. Immediate next steps include independent probe/conjugate characterization (efficiency and topological charge), phase-resolved OAM diagnostics via heterodyne interferometry, and optimization of operating points (detuning, temperature, and pump power) to maximize conversion efficiency while minimizing absorption. Together, this establishes hot-vapor double-Λ FWM as a versatile, detuning-tunable engine for generating and manipulating optical vortex beams.

## Acknowledgments

We acknowledge financial support from the Israel Innovation Authority. We thank Dr. Gilad Orr for insightful discussions.

## Data availability

The data are not publicly available. The data are available from the authors upon reasonable request.

## Competing Interests

The authors declare no competing interests.